\documentclass[aps,prl,twocolumn,showpacs]{revtex4}
\usepackage{amsmath}
\usepackage{epsfig}
\usepackage{amssymb}
\usepackage{dcolumn}
\usepackage{graphicx}
\usepackage{dcolumn}

\begin{document}
\date{\today}

\title{Nonlinear Supression of Tunneling and Strong Localization in Bose-Einstein Condensates with Spatially Inhomogeneous Interactions}

\author{V\'{\i}ctor  M. P\'erez-Garc\'{\i}a}
\email{victor.perezgarcia@uclm.es}
\homepage{http://matematicas.uclm.es/nlwaves}

\affiliation{Departamento de Matem\'aticas, E. T.
S. de Ingenieros Industriales, and Instituto de Matem\'atica Aplicada a la Ciencia y la Ingenier\'{\i}a (IMACI),
Universidad de Castilla-La Mancha, 13071 Ciudad Real, Spain.}

\begin{abstract}
We study the properties of the ground state of Bose-Einstein condensates with spatially inhomogeneous
interactions and show that the atom density experiences a strong localization at the spatial region where the scattering length is close to zero while tunneling to regions with positive values of the scattering length is strongly supressed by the nonlinear interactions.
\end{abstract}

\pacs{03.75.Lm, 05.45.Yv, 42.65.Tg}

\maketitle

The experimental generation of Bose-Einstein condensates (BEC) with ultracold dilute atomic
vapors \cite{1} has turned out to be of exceptional importance for physics. 
The formation of a condensate occurs when the temperature is low enough and most of the atoms occupy the ground state of the system. This process is visible both in momentum and in real space due to the spatial inhomogeneities exhibited by the order parameter on a macroscopic scale because of the trapping potentials.

The properties of the ground state of trapped BECs are well known. In the mean field limit
simple analytical expressions are available  in the  Thomas-Fermi approximation \cite{Thomas-Fermi} and beyond \cite{Dalfovo}. These approximations and direct numerical simulations describe accurately the properties of the experimentally found ground states \cite{Numerics}.
 
Nonlinear interactions between atoms in a Bose-Einstein condensate are dominated by the two-body collisions that can be controlled by the so-called Feschbach resonance (FR) management \cite{FB1}. The control in time of the scattering length has been used to generate bright solitons \cite{bright} and induce collapse \cite{collapse} and has been the basis for theoretical proposals to obtain different nonlinear waves \cite{Kono1,Kono2,Ueda}. Moreover, interactions can be made spatially dependent by acting on either the magnetic field or the laser intensity (in the case of optical control of FRs \cite{FB2}) acting on the Feschbach resonances. This possibility has motivated many theoretical studies on the behavior of solitons in Bose-Einstein condensates (BECs) with spatially inhomogeneous interactions \cite{Victor1,Garnier,Panos,Boris2,tuti}. 
 
  In this paper we study the effect of specific spatially inhomogeneous interactions on the ground state of a BEC in the mean field limit. We will show that when the scattering length is non-negative, a striking localization phenomenon of the atom density occurs at the regions where the scattering length vanishes. By tuning appropriately the control (magnetic or optical) fields this phenomenon can be used to design regions with large particle densities and prescribed geometries. Another interesting phenomenon to be studied in this letter is the nonlinear limitation of tunneling of atoms to the regions in which the interactions are stronger.
    
 \emph{The model and its Thomas-Fermi limit.-} The ground state of  a BEC in the mean field limit is the real, positive solution of the Gross-Pitaevskii equation 
 \begin{equation}\label{gs}
 \lambda \phi = -\frac{1}{2} \Delta \phi + V(\boldsymbol{x}) \phi + g(\boldsymbol{x}) |\phi|^2 \phi,
 \end{equation}
 which minimizes the energy 
 \begin{equation}
 E(\phi) =  \int_{\mathbb{R}^3} \left[  \frac{1}{2} \left|\nabla \phi\right|^2 + V(\boldsymbol{x}) \left|\phi\right|^2 + \frac{1}{2} g(\boldsymbol{x}) |\phi|^4\right],
 \end{equation}
under the constraint of a fixed number of particles $N = \int_{\mathbb{R}^3} |\phi|^2$.
Eq. (\ref{gs}) is written in nondimensional units where the coordinates $\boldsymbol{x}$ and time $t$ are measured in units of  $a_0=\sqrt{\hbar/m\omega}$ and $1/\omega$, respectively, while the energies and frequencies are measured in  units of $\hbar\omega$ and $\omega$ respectively,  $\omega$ being  a characteristic frequency of the potential. Finally, $g(\boldsymbol{x}) =4\pi a(\boldsymbol{x})/a_0$ is proportional to the local value of the s-wave
scattering length $a$.
 
The Thomas-Fermi approximation proceeds by neglecting the kinetic energy or equivalently the term proportional to $\Delta \phi$ in Eq. (\ref{gs}). In the case of spatially homogeneous interactions $g(\boldsymbol{x}) = g_0$ this leads to $\phi_{TF}(\boldsymbol{x}) = \left[\left(\lambda-V(\boldsymbol{x})\right)/g_0\right]^{1/2}$. When the interactions are spatially inhomogeneous the same formal manipulation leads to 
\begin{equation}
\phi_{TF}(\boldsymbol{x}) = \sqrt{\left(\lambda-V(\boldsymbol{x})\right)/g(\boldsymbol{x})},
\end{equation}
which diverges on the set $G = \{\boldsymbol{x} \in \mathbb{R}^3:  g(\boldsymbol{x}) = 0\}$. Obviously, in the vicinity of  $G$ the Thomas-Fermi approximation is not correct but anyway this divergence is a first indication of a phenomenon to be studied in this paper: \emph{the tendency of the ground state in Bose-Einstein condensates with spatially inhomogeneous interactions to localize strongly on the regions where the scattering length is close to zero provided the system is sufficiently nonlinear}, i.e. for sufficiently large values of  $gN$.

 \emph{A simple example.-} Let us first consider an exactly solvable  ``toy" example which displays the main features of the phenomena to be studied in this paper: a quasi-one dimensional BEC in a box, i.e. setting $V(x) = 0$,  $\phi(x = \pm L) = 0$ with scattering length given by 
 $g(x) = g_0$, for $|x| < a$ and $g(x) =  0,$ for  $|x| >a$.
 
 In this simple case, Eq. (\ref{gs})  becomes
 \begin{subequations} \label{gssimple}
 \begin{eqnarray}
\phi_{xx} + 2 \lambda \phi  =  & 2 g_0  \phi^3, & |x| \leq a, \\
 \phi_{xx} + 2 \lambda \phi =  & 0, & a <|x| < L,
 \end{eqnarray}
 \end{subequations}
 and its positive, even solution, satisfying the boundary conditions
$\phi(\pm L) = 0, \phi'(0) = 0$
 is given by  
\begin{equation}\label{pepa}
 \phi(x)   =   C \sin \left[\sqrt{2\lambda} \left(x-L\right)\right], a <|x| < L,
 \end{equation}
while for $|x|<a$ the solutions are given by
 \begin{equation}
  \phi(x)    = \begin{cases} \sqrt{\frac{\alpha \lambda}{g_0}} \ \text{sn}\left(x\sqrt{\lambda \alpha}+\delta; k^2\right),   & \lambda < \lambda_*\\
 \pi/\left(2\sqrt{2g_0}|a-L|\right), & \lambda = \lambda_*, \\
 \sqrt{\frac{\alpha \lambda}{g_0}} \ \text{dc}\left(x\sqrt{\lambda \alpha}; k^2\right), &  \lambda > \lambda_*
  \end{cases}
   \end{equation}
   where $\alpha(k) = 2/(1+k^2)$, sn and dc are two of the standard Jacobi elliptic functions and $k$ is the elliptic modulus. Both the elliptic modulus and amplitude $C$ can be obtained from the matching conditions for $\phi(a)$ and $\phi'(a)$. These conditions also give $\lambda_* = \pi^2/\left[8(a-L)^2\right]$. Finally, the cutoff value of the chemical potential $\lambda_c$ can be obtained from the condition of maximum slope at $x=a$ which leads to $\lambda_c = \pi^2/[2(a-L)^2]$. In that case the number of particles in the condensate is infinite, since the amplitude in the outer region $C \rightarrow \infty$. 
    \begin{figure}
 \epsfig{file=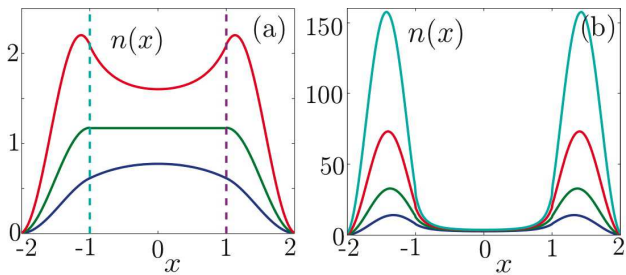,width=0.95\columnwidth}
 \caption{[Color online] Spatial distribution of the density $n(x) = |\phi(x)|^2$ for the ground state solutions of Eq. (\ref{gssimple}) with $L=2, a =1$ and different values of the chemical potential $\lambda$ corresponding to different values of $N$ (a) From the lower to the upper curve $g_0N =2, 3.55$ (corresponding to $\lambda_*$) and $g_0N=6$. (b) From the lower to the upper curve $g_0N=25, 50, 100, 200$.  \label{prima}}
 \end{figure}
 
The  spatial profiles of the ground state density for different values of $g_0N$ shown in Fig. \ref{prima} support our conjecture based on the Thomas-Fermi solution, i.e. the existence of a strong localization of the atom density in the region where the interactions vanish.  

It is also remarkable that the atom density in the inner part of the domain, i.e. the region where there are nonlinear interactions, remains almost constant independently on the number of particles in the condensate once a certain critical density is achieved. This region corresponds to the nonlinear analogue of the classically forbidden region in ordinary potentials and is energetically less favourable due to the extra repulsive energy provided by the nonlinear interactions. However, the tunneling of atoms in this region is essentially limited to a constant value, \emph{independently of the number of atoms}, which differs essentially from ordinary tunneling. 
   
\begin{figure}
 \epsfig{file=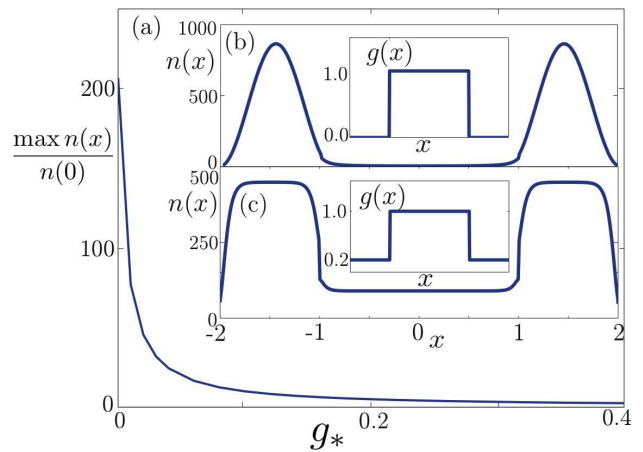,width=0.95\columnwidth}
 \caption{[Color online]  (a) Ratio between the maximum atom density and the atom density at $x=0$ as a function of the scaled scattering length $g_*$ on the spatial region $|x|<a$ (for $g_0 = 1$) (b-c) Scaled atom density profiles for (b) $g_*=0$ and (c) $g_* = 0.2$ in both cases for a total scaled number of particles $N=1000$. The insets show the profile of $g(x)$. \label{nueva}}
 \end{figure}

The supression of tunneling depends strongly on the value of the scattering length in the inner region, that we have taken to be zero up to now. If instead we set $g(x) = g_*$ when $|x|<a$ and study the dependence of the ratio between the maximum atom density and the atom density at $x=0$ (which is a measure of the amplitude of the tunneling), we find a strong dependence on this parameter as shown in Fig. \ref{nueva}(a).

This effect is also seen in the atom density profiles when comparing the cases with $g_*=0$ [Fig. \ref{nueva}(b)] and $g_*=0.2$ [Fig. \ref{nueva}(c)] for $N=1000$. Larger values of $N$ lead to a stronger effect.
   
\begin{figure}
 \epsfig{file=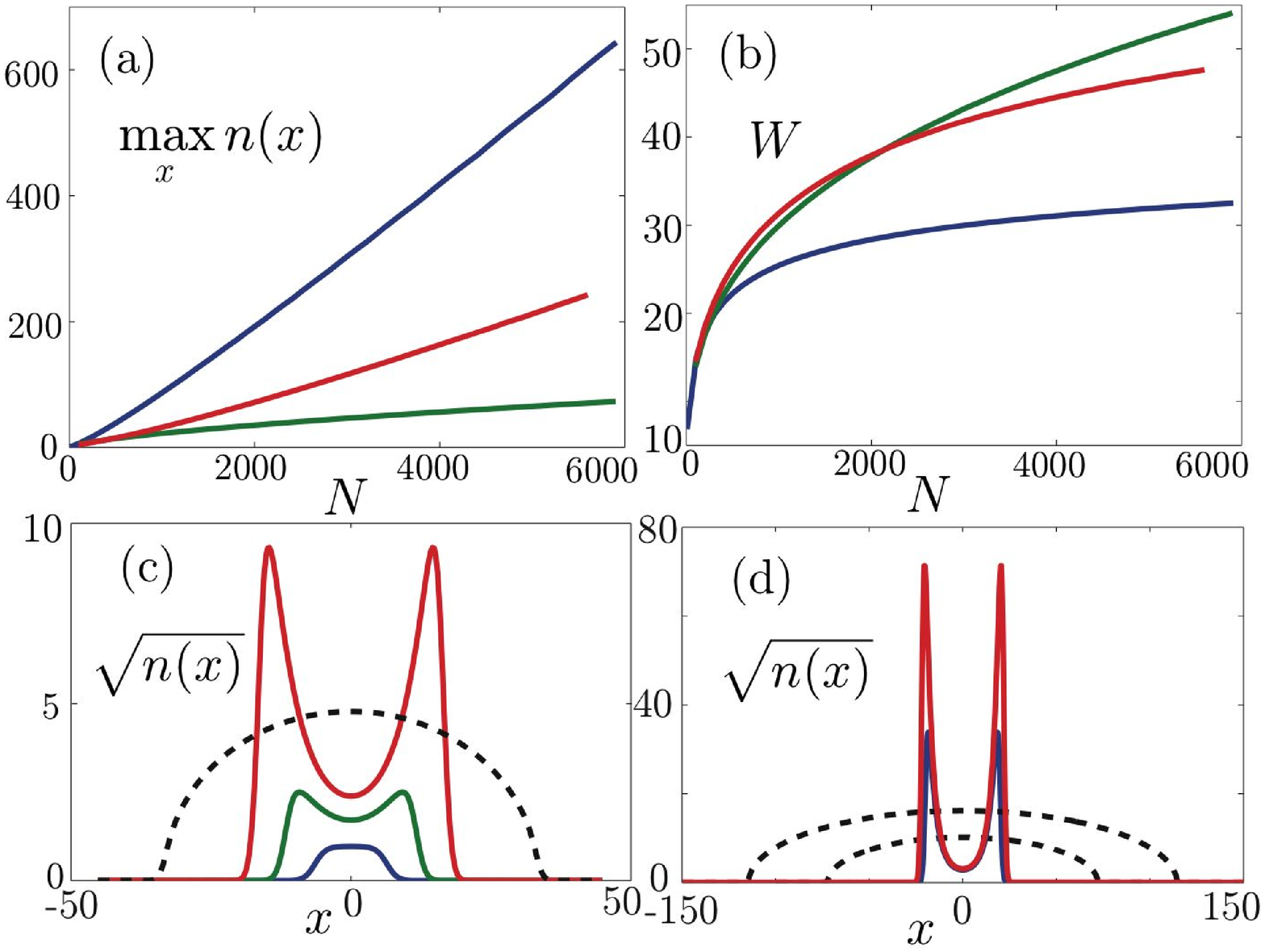,width=\columnwidth}
 \caption{[Color online] Ground states of Eq. (\ref{gs}) for $V(x) = 0.02 x^2$ and $g_0(x) = 1, g_1(x)=\exp (-x^2/200)$, and $g_2(x) = \exp(-x^2/50)$,  for different values of the scaled number of particles $N$. (a) Maximum particle density $\max_x n(x)$ and (b) width $W^2 = \int x^2 n(x)dx/N$ for $g_0$ (green), $g_1(x)$ (red) and $g_2(x)$ (blue). (c) Spatial profiles of $\sqrt{n(x)}$ for $N=10$ (blue), $N=100$ (green), $N=1000$ (red) for $g(x)=g_1(x)$. The dashed black line is the ground state with homogeneous interactions and $N=1000$. (d) 
Same as (c) but for $N=10000$ (blue), and $N=40000$ (green),  in comparison with the case of spatially homogeneous interactions (dashed black lines). \label{segunda}}
 \end{figure}

 \emph{Numerical results.-} Let us now consider Eq. (\ref{gs}) in more realistic quasi one-dimensional scenarios by including a potential $V(x) = 0.02 x^2$.  We have computed numerically the ground state for  scattering lengths of the form
 $g_0(x) = 1, g_1(x) = \exp(-x^2/200), g_2(x) = \exp(-x^2/50)$. These choices allow us to study different degrees of localization of the interactions starting from the case of no localization.   Our results are summarized in Fig. \ref{segunda}. 
 
In Fig. \ref{segunda}(a) we observe how the maximum density ($n(x) = |\phi(x)|^2$)  increases drastically for spatially decaying nonlinearities (blue and red curves) as a function of the number of effective number of particles in the quasi-one dimensional condensate $N$. This amplitude growth is due to a strong localization effect near the region where $g(x)$ vanishes as it is seen in Fig. \ref{segunda}(c,d). In contrast, the condensate density for spatially homogeneous interactions grows slowly according to the Thomas-Fermi  prediction $\max n(x) \propto N^{2/3}$.  When the number of particles is small, the size of the atomic cloud  is smaller than the  localization region of $g(x)$. For larger values of $N$ the ground state extends beyond the localization region of $g(x)$ and the atom density becomes more and more localized near its edge. This effect is more clear for larger number of particles and is accompanied by a saturation in the amplitude growth in the region where $g(x)$ is far from zero [Fig. \ref{segunda}(d)]. 

In the case of spatially homogeneous interactions the width grows according to the law $W \propto N^{1/3}$  [Fig. \ref{segunda}(b)]. When interactions are spatially dependent and due to the localization of the amplitude close to the zero of $g(x)$ the width growth saturates for large values of $N$ to a value depending on the size of $g(x)$.

\begin{figure}
 \epsfig{file=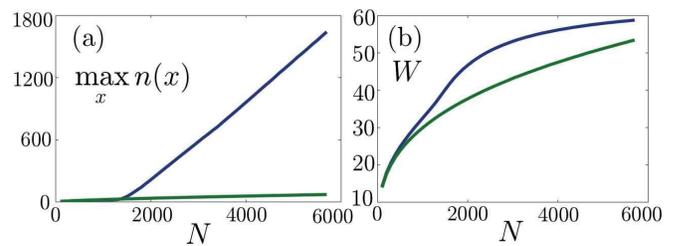,width=\columnwidth}
 \caption{[Color online] Ground state maximum particle density (a) and width $W = \int x^2 n(x)dx/N$ (b) for $V(x) = 0.02 x^2$ and $g_0(x) = 1$ (red lines) and $g_4(x)=(1-0.001 x^2)_+$ (blue lines). \label{tertia}}
 \end{figure}

These effects are even more clear when the nonlinearity decays to zero faster. For instance, taking a nonlinear coefficient given by $g_4(x) = (1-0.001 x^2)_+$ (i.e. an inverted parabola with maximum amplitude $g=1$ at $x=0$ and zero values for $|x|>31.6$) as shown in Fig. \ref{tertia}, we see that the maximum densities are even higher than for $g_1(x), g_2(x)$. In this case, in comparison with our first simple example, we can see an even stronger localization since the existence of the potential makes energetically more favourable the localization close to the point where the interactions vanish.

   \begin{figure}
 \epsfig{file=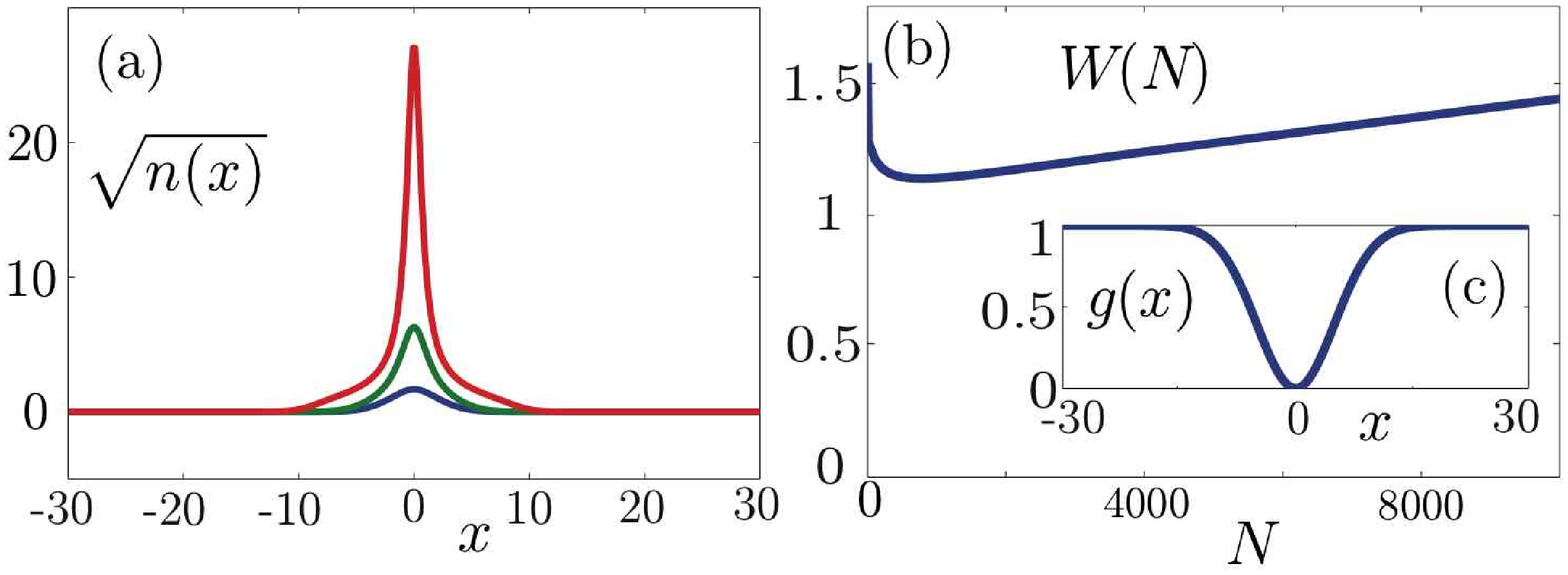,width=\columnwidth}
 \caption{[Color online] Ground states for $V(x) = 0.02 x^2$ and $g(x) = 1-\exp(-x^2/50)$ (shown in panel (c)). (a) $\sqrt{n(x)}$  for $N=10$ (blue line), $N=100$ (green line) and $N=1000$ (red line). (b) Condensate width as a function of $N$. \label{cuarta}}
 \end{figure}

 Finally, in Fig. \ref{cuarta} we get again the localization phenomenon but now near $x=0$ for localized interactions given by $g(x) = 1- \exp(-x^2/50)$ and $V(x) = 0.02 x^2$. It is interesting  how localized the density becomes to ``avoid" penetrating into the regions with appreciable values of the scattering length.
 

 
 \emph{Rigorous results.-} In Ref. \cite{Lopez} the equation
 \begin{equation}\label{delta}
 - \Delta u = \lambda u - a(x) \left(u^r+f(x,u)\right)u,
 \end{equation}
 was considered on a bounded domain  $\Omega \subset \mathbb{R}^N$ with $C^{1,1}$ boundary under the conditions $u|_{\partial \Omega} = 0$. $r>0$, 
 $a(x) \in C(\bar{\Omega})$ is nonnegative and the open set $D = \left\{ x \in \Omega: a(x) >0\right\}$ satisfies $\bar{D} \subset \Omega$ and possesses a finite number of $C^1$ connected components 
 $D_j, 1 \leq j \leq l$ such that $\bar{D}_i \cap \bar{D}_j = \emptyset$ if $i \neq j$. The function $f: \bar{\Omega} \times [0, +\infty) \rightarrow \mathbb{R}$, satisfies $f, f_u = \partial f/\partial u \in C\left(\bar{\Omega} \times [0, \infty); \mathbb{R}\right)$ and the growth conditions $
 \lim_{u\downarrow 0} f(\cdot,u) = 0, \lim_{u\uparrow \infty} f_u(\cdot, u)/u^{r-1} = 0,$
 uniformly in $\bar{\Omega}$.
 In the case when $f = 0$, Eq. (\ref{delta}) is the logistic model of population dynamics \cite{16} where $\Omega$ is the region inhabited by the species $u$, $\lambda$ measures its birth rate and $a(x)$ measures the capacity of $\Omega$ to support the species $u$. Given an open subset $\Omega_1 \subset \Omega$ with a finite number of components let us denote by
 $\sigma_1^{\Omega_1}[-\Delta]$ the minimum of the principal eigenvalues of the operator $-\Delta$ on each of the components of $\Omega_1$ subject to homogeneous Dirichlet boundary conditions on $\partial \Omega_1$ it was proven \cite{Lopez} that Eq. (\ref{delta}) has an unique positive solution for $\lambda \in (\sigma_1^{\Omega}[-\Delta], \sigma_1^{\Omega_0}[-\Delta])$ and that no such solution exists for $\lambda \geq \sigma_1^{\Omega_0}[-\Delta]$. Moreover, it grows to infinity when $\lambda \uparrow \sigma_1^{\Omega_0}[-\Delta]$ on compact subsets of  
 $\Omega_0$ and stabilize in $D$ to the minimal weak solution of Eq. (\ref{delta}) with singular boundary conditions $u|_{\partial D} = \infty$.

From Eq. (\ref{delta}) we can obtain our quantum mechanical problem described by Eq. (\ref{gs}) by choosing $f(x,u) = 0, a(x) = 2 g(x), r=2$. However, the results of Ref. \cite{Lopez} essentially require a bounded domain and a continuous $g(\boldsymbol{x})$
and thus are not directly applicable to the examples discussed in this paper. However, they support 
 the localization phenomenon and the saturation of the tunneling amplitude in formally related problems.

 \emph{Conclusions and discussion.-} We have studied the ground state of Bose-Einstein condensates with repulsive spatially varying interactions and shown that 
 the atom density localizes dramatically near the regions where the interactions are close to zero when the nonlinear interactions are relevant enough. This behavior  does not depend on the spatial dimensionality and can be used to design very effectively one, two or three-dimensional spatial distributions with large values of the atom density by acting on the control field, e.g.
 using micromagnets to change locally the magnetic field or localized laser beams. This behavior is not so easy to achieve with ordinary potentials acting on the condensate.
Our results could also be useful for atom lithography, atom beam guiding  or other applications where a precise control of the positioning of large values of the density of an atomic cloud is required.

We have also found a limitation of tunneling to the regions with larger values of the scattering length which is independent on the number of particles and differs essentially from classical tunneling in ordinary potentials. 
 
These phenomena  may appear in multicomponent condensates that 
  offer a much wider range of possibilities for their ground states depending on the type of components and their relative interactions  \cite{ground}. It would be interesting to study the situation in which the variations of the scattering length make the components to be miscible or inmiscible in different spatial regions and  what is the geometry of the resulting domains.

These phenomena could also be observed in nonlinear optical systems. Although ordinary optical materials have small nonlinearities, there are media with a giant nonlinear response \cite{EIT} as it happens when a probe laser beam propagates in a medium with  transparency induced electromagnetically by a second coupling field. Therefore, managing  the parameters it could be posible to find an optical version of the phenomena presented here.
 
\acknowledgements

I want to acknowledge R. Pardo (U. Complutense) for discussions. This work has been partially supported by grants FIS2006-04190
 (Ministerio de Educaci\'on y Ciencia, Spain) and PAI-05-001 (Junta de Comunidades de Castilla-La Mancha, Spain).

  \end{document}